\documentclass{Interspeech}
\usepackage{multirow}
\usepackage{kotex} 
\usepackage{threeparttable}
\usepackage{pifont}
\usepackage{algorithm}
\usepackage{algpseudocode}
\usepackage{algorithmicx}
\usepackage{float}

\interspeechcameraready

\title{Naturalness-Aware Curriculum Learning with Dynamic Temperature for Speech Deepfake Detection}

\author[affiliation={1}]{Taewoo}{Kim}
\author[affiliation={1}]{Guisik}{Kim}
\author[affiliation={1}]{Choongsang}{Cho}
\author[affiliation={1}]{Young Han}{Lee}

\affiliation{}{Korea Electronics Technology Institute}{South Korea}

\email{kimtaewoo@keti.re.kr, specialre@naver.com, ideafisher@keti.re.kr, yhlee@keti.re.kr}

\usepackage{comment}

\begin{document}

\maketitle

\begin{abstract}
Recent advances in speech deepfake detection (SDD) have significantly improved artifacts-based detection in spoofed speech. However, most models overlook speech naturalness, a crucial cue for distinguishing bona fide speech from spoofed speech. This study proposes naturalness-aware curriculum learning, a novel training framework that leverages speech naturalness to enhance the robustness and generalization of SDD. This approach measures sample difficulty using both ground-truth labels and mean opinion scores, and adjusts the training schedule to progressively introduce more challenging samples. To further improve generalization, a dynamic temperature scaling method based on speech naturalness is incorporated into the training process. A 23\% relative reduction in the EER was achieved in the experiments on the ASVspoof 2021 DF dataset, without modifying the model architecture. Ablation studies confirmed the effectiveness of naturalness-aware training strategies for SDD tasks.
\end{abstract}

\section{Introduction}
Advancements in speech synthesis technologies, such as text-to-speech (TTS) and voice conversion (VC), have enabled various applications in virtual assistants, entertainment, and accessibility. However, the increasing realism of synthetic speech has raised significant societal concerns, such as financial fraud, impersonation attacks, and misinformation dissemination. As these risks grow, robust speech deepfake detection (SDD) techniques are crucial to counter these threats and protect voice-based systems. In this context, anti-spoofing challenges such as ASVspoof~\cite{todisco19_interspeech, yamagishi21_asvspoof} have become key benchmarks, significantly contributing to the development of robust and effective SDD approaches.

Previous SDD approaches primarily focused on analyzing spoofing artifacts in synthetic speech and leveraging various acoustic characteristics for detection. Early approaches~\cite{ge21_asvspoof,li2021replay,jung2022aasist} successfully identified spoofing artifacts by extracting handcrafted features such as linear frequency cepstral coefficients, constant-Q transform, and sinc convolution across different frequency bandwidths. In addition, some studies~\cite{kim2023phase,yang2021modified} explored phase distortions in speech, thereby capturing inconsistencies in temporal and spectral characteristics. Although effective in specific scenarios, the generalizability of such models is limited when novel attacks or unseen environments are encountered.

To improve the generalization performance of SDD systems, recent studies~\cite{tak22_odyssey, guo2024audio, rosello23_interspeech, truong24b_interspeech} have adopted self-supervised learning (SSL) models pretrained on large-scale speech corpora as feature extractors, such as XLS-R~\cite{babu22_interspeech} and WavLM~\cite{chen2022wavlm}. These models learn rich acoustic representations to detect artifacts that are often missed by handcrafted features. However, recent advances in vocoder technology have led to nearly artifact-free outputs \cite{bak2023avocodo}, making it increasingly difficult to distinguish synthetic speech from real recordings. These improvements have been incorporated into state-of-the-art TTS systems~\cite{popov2021grad, mehta2024matcha}, enabling them to generate high-fidelity speech with minimal distortion and rendering artifacts-based detection methods increasingly ineffective. Nonetheless, a perceptual gap remains between synthetic and human speech, particularly in terms of naturalness~\cite{minixhofer23_interspeech}. Thus, incorporating perceptual cues such as speech naturalness is essential for enhancing the robustness and generalization of SDD systems.

In this study, we propose naturalness-aware curriculum learning for SDD, leveraging mean opinion score (MOS) predictions to enhance generalization. Our approach follows a training strategy that involves starting with relatively easy samples and progressively introducing more challenging ones, helping the model adapt effectively. In addition, we introduce a dynamic temperature scaling method that adjusts the softmax temperature based on sample difficulty during training. This method helps regulate model confidence, preventing overfitting and enhancing robustness. Furthermore, we employ an SSL-based baseline model to obtain rich acoustic representations, demonstrating that our approach effectively captures the naturalness of synthetic speech and improves SDD performance. In the experiments, we evaluate the efficacy of the proposed approach by applying it to various SDD models. Moreover, we assess the generalizability of the proposed approach across multiple datasets and performed ablation studies to further substantiate its benefits.

\section{Related work}
\subsection{Curriculum learning for neural networks}
Curriculum learning (CL), first introduced by Bengio et al.~\cite{bengio2009curriculum}, improves model performance by gradually increasing the difficulty of the training samples. Inspired by the way humans learn by beginning with easy concepts and gradually moving to more challenging ones, CL improves the generalizability of neural networks. CL has been widely explored in machine learning, where the sample difficulty is typically determined by factors such as data noise, variability, and training loss.

Recently, Song et al.~\cite{song2024towards} applied CL to face deepfake detection, dynamically adjusting sample difficulty based on instance loss and facial quality, leading to improved performance. Building on this idea, we apply CL to speech deepfake detection (SDD) by leveraging mean opinion scores (MOS) to assess speech naturalness. To the best of our knowledge, this is the first study to apply CL to SDD.

\subsection{Temperature scaling}
Temperature scaling was initially introduced for knowledge distillation~\cite{hinton2015distilling} and has also been applied to confidence calibration~\cite{guo2017calibration} in classification models. Recently, Dabre et al. \cite{dabre2021investigating} demonstrated that applying temperature scaling during training can reduce overfitting and improve generalization in neural machine translation. Similarly, Khaertdinov et al.~\cite{khaertdinov2022dynamic} explored dynamic temperature scaling based on instance similarities within a contrastive learning framework to mitigate the impact of false negatives and improve representation learning.

Inspired by these studies, we introduce dynamic temperature scaling for SDD, where the temperature is dynamically adjusted based on sample difficulty during training. This allows the model to increase confidence for easier samples and decrease it for harder ones, preventing overfitting and improving generalization.
\vspace{-0.3cm}

\section{Method}
\begin{figure}[t]
  \centering
  \includegraphics[width=\linewidth]{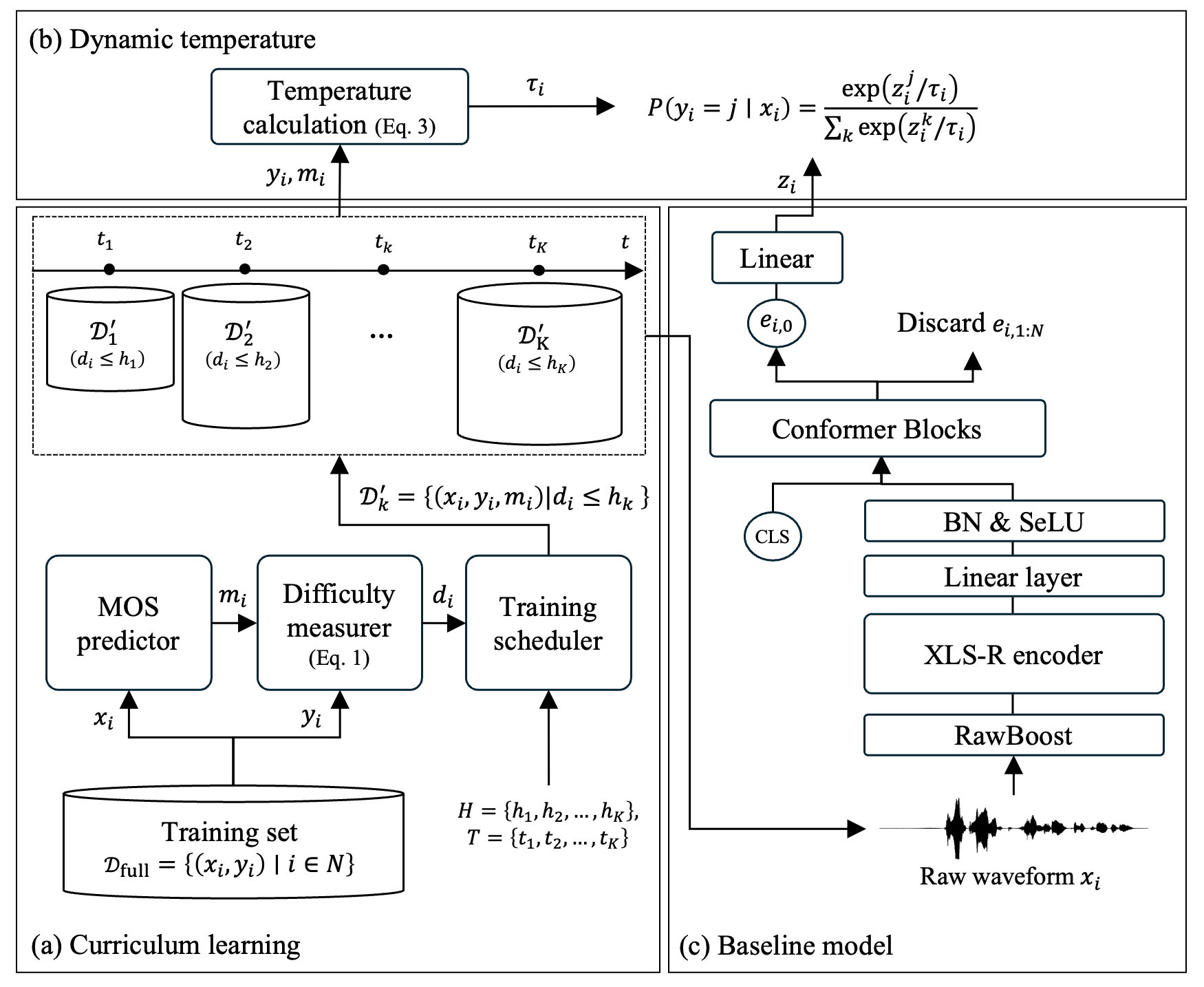}
  \vspace{-15pt}
  \caption{Overview of the proposed training framework.}
  \label{fig:overview}
  \vspace{-0.5cm}
\end{figure}

\subsection{Overview}
Figure~\ref{fig:overview} presents an overview of the proposed training framework, which integrates curriculum learning and dynamic temperature scaling to enhance the speech deepfake detection (SDD) model. In Figure~\ref{fig:overview}(a), the curriculum learning component organizes training by measuring sample difficulty and adjusting the training schedule accordingly. It consists of three key modules: MOS predictor, difficulty measurer, and training scheduler. The MOS predictor estimates the perceptual naturalness of speech samples, whereas the difficulty measurer computes difficulty scores based on MOS predictions and label information. Using these scores, the training scheduler gradually expands the dataset by incorporating increasingly difficult samples, thereby ensuring a structured learning process.

Figure~\ref{fig:overview}(b) illustrates that the dynamic temperature scaling method adjusts the softmax confidence during training, ensuring it varies according to sample difficulty. This allows the model to assign higher confidence to easier samples while focusing more on perceptually challenging ones. By integrating these components, the proposed framework enhances both learning efficiency and model robustness, ensuring that training adapts dynamically to sample difficulty.

\subsection{Curriculum learning}
\subsubsection{MOS-based difficulty measurer}
To construct an effective training curriculum, it is essential to quantify the difficulty of each sample. We use the MOS, a widely used metric for assessing speech quality, to measure the difficulty of the training data. The MOS quantifies the perceptual quality of speech, where higher values correspond to more natural and human-like speech. 

As illustrated in Figure~\ref{fig:overview}(a), we first employ an MOS predictor to estimate the naturalness MOS \(m_i\) for each speech sample \(x_i\), where \(i\) denotes the index of the training sample. The MOS is then normalized to the range \([0,1]\) to represent the difficulty on a consistent scale, resulting in \( \hat{m}_i \). Based on the normalized MOS \( \hat{m}_i \) and sample labels \( y_i \), the difficulty score \( d_i \) is defined as follows:
\begin{equation}
    d_i =
    \begin{cases}
        \hat{m}_i, & \quad \text{if } y_i = 0 \text{ (spoof)}, \\[2pt]
        1 - \hat{m}_i, & \quad \text{if } y_i = 1 \text{ (bona fide)}. 
    \end{cases}
    \label{eq:difficulty}
\end{equation}
This formulation ensures that deepfake samples with higher MOS are assigned higher difficulty scores, whereas bona fide samples with higher MOS are considered easier due to their perceptual clarity. 

\subsubsection{Training scheduler}
The proposed training schedule follows an easy-to-hard CL approach that progressively introduces training samples based on their difficulty. First, the training samples are sorted according to their difficulty scores, which are computed using Equation~\ref{eq:difficulty}. The training schedule is then controlled by two hyperparameters: difficulty levels \( H = \{ h_k \}^{K}_{k=1} \)  and pacing sequence \( T = \{ t_k \}^{K}_{k=1} \), where \( K \) denotes the total number of curriculum stages and \( k \) represents the stage index.

The difficulty levels \( H \) consist of a set of difficulty scores that determine the composition of the training subsets. To enable stepwise learning, multiple subsets are predefined, where each subset contains samples with difficulty scores lower than the corresponding level \( h_k \). The pacing sequence \( T \) specifies the epoch \( t_k \) at which these training subsets are introduced into the existing training data. 

Initially, training begins with the easiest subset, and as training progresses, expanded subsets are introduced at the epochs specified in \( T \). This stepwise expansion guides the model to gradually adapt to increasingly complex data distributions.

\begin{table*}[t]
\caption{Evaluation of the proposed curriculum learning and dynamic temperature on the ASVspoof 2021 dataset.}

\centering
\begin{tabular}{l|cc|cc|c|cc|c}
\hline
\multicolumn{1}{c|}{\multirow{2}{*}{\textbf{Model}}} & 
\multicolumn{2}{c|}{\textbf{Train}} & 
\multicolumn{2}{c|}{\textbf{LA (Fix)}} & 
\textbf{DF (Fix)} & 
\multicolumn{2}{c|}{\textbf{LA (Var)}} & 
\textbf{DF (Var)} \\ \cline{2-9}
\multicolumn{1}{c|}{} &
CL & DT & EER(\%) & min t-DCF & EER(\%) & EER(\%) & min t-DCF & EER(\%) \\ \hline

\multirow{4}{*}{XLS-R AASIST}
& \ding{55} & \ding{55} & \textbf{1.00} & \textbf{0.211} & 2.36 & \textbf{0.89} & \textbf{0.209} & 3.31 \\ 
& \ding{51} & \ding{55} & 1.05 & 0.214 & \underline{2.27} & 1.12 & 0.214 & \underline{3.00} \\ 
& \ding{55} & \ding{51} & \underline{1.01} & \underline{0.213} & 2.37 & 1.17 & 0.215 & 3.17 \\ 
& \ding{51} & \ding{51} & 1.03 & 0.213 & \textbf{2.09} & \underline{1.06} & \underline{0.214} & \textbf{2.52} \\ \hline

\multirow{4}{*}{XLS-R Transformer}
& \ding{55} & \ding{55} & 1.55 & 0.227 & 2.48 & 1.48 & 0.226 & 2.84 \\
& \ding{51} & \ding{55} & \textbf{1.16} & \textbf{0.217} & 2.36 & \underline{1.24} & \underline{0.219} & 2.83 \\ 
& \ding{55} & \ding{51} & 1.26 & 0.220 & \underline{2.13} & 1.29 & 0.220 & \underline{2.68} \\ 
& \ding{51} & \ding{51} & \underline{1.17} & \textbf{0.217} & \textbf{2.00} & \textbf{1.15} & \textbf{0.217} & \textbf{2.48} \\ \hline

\multirow{4}{*}{XLS-R Conformer}
& \ding{55} & \ding{55} & 1.09 & 0.215 & 2.45 & 1.03 & 0.213 & 2.72 \\
& \ding{51} & \ding{55} & \underline{0.92} & \underline{0.210} & \underline{2.11} & \underline{0.76} & \underline{0.206} & \underline{2.44} \\ 
& \ding{55} & \ding{51} & 1.23 & 0.219 & 2.23 & 1.03 & 0.214 & 2.57 \\ 
& \ding{51} & \ding{51} & \textbf{0.89} & \textbf{0.208} & \textbf{1.88} & \textbf{0.69} & \textbf{0.204} & \textbf{2.21} \\ \hline
\end{tabular}
\label{tab:proposed_performance}
\vspace{-0.4cm}
\end{table*}

\vspace{-0.2cm}
\subsection{Dynamic temperature}
To further enhance the robustness of SDD models, we introduce a dynamic temperature scaling method. This approach adjusts the temperature parameter  \( \tau \) based on the normalized MOS \( \hat{m} \) of each sample, allowing the model to account for perceptual difficulty in classification.

First, we perform a grid search over the MOS values on the training set to obtain the normalized MOS threshold \( \hat{m}_{\text{th}} \) that best distinguishes between bona fide and spoofed speech. Here, \(\hat{m}_{\text{min}} \) and \(\hat{m}_{\text{max}} \) denote the minimum and maximum normalized MOS, respectively. We introduce the asymmetry factor \( \lambda \) to handle different difficulty scaling on both sides of the threshold. The asymmetry factor \( \lambda \) is computed as follows:
\begin{equation}
    \lambda = \frac{\hat{m}_{\text{max}} - \hat{m}_{\text{th}}}{\hat{m}_{\text{th}} - \hat{m}_{\text{min}}}.
    \label{eq:asymmetry_factor}
\end{equation}
Using this factor \( \lambda \), the temperature \( \tau_i \) with normalized MOS \( \hat{m}_i \) is defined as:
\begin{equation}
    \tau_i =
    \begin{cases}
        1 + \lambda (\hat{m}_i - \hat{m}_{\text{th}}), & \quad \text{if } y_i = 0 \text{ (spoof)}, \\[2pt]
        1 - \lambda^{-1} (\hat{m}_i - \hat{m}_{\text{th}}), & \quad \text{if } y_i = 1 \text{ (bona fide)}.
    \end{cases}
    \label{eq:temperature}
\end{equation}
During training, the final logits are scaled by the temperature \( \tau_i \), resulting in the following modified softmax operation:
\begin{equation} 
P(y_i = j \mid x_i) = \frac{\exp(z_i^j / \tau_i)}{\sum_{k} \exp(z_i^k / \tau_i)},
\label{eq:softmax_temperature} 
\end{equation}
where \( z_i^j \) denotes the logit for class \( j \) of sample \( i \), and \( \tau_i \) is dynamically adjusted according to the perceptual difficulty of the sample. This scaling mechanism regulates confidence in the softmax function, reducing overconfidence in predictions for more challenging samples while increasing confidence for easier ones during training.

\section{Experiments}
\subsection{Datasets and metrics}
In the experiments, all models were trained on the ASVspoof 2019 logical access (LA) dataset~\cite{todisco19_interspeech}. The training set included 2,580 bona fide and 22,800 spoofed utterances, while the validation set included 1,064 bona fide and 22,296 spoofed utterances, which were generated using four TTS and two VC algorithms. We computed the mean opinion score (MOS) of the training set using UTMOS\footnote{\url{https://github.com/sarulab-speech/UTMOS22}}~\cite{saeki22c_interspeech} to assess speech naturalness. To evaluate the performance of the proposed methods, we used the evaluation subsets of the ASVspoof 2021 LA and DeepFake (DF) datasets~\cite{yamagishi21_asvspoof}. The LA dataset consisted of 148,176 utterances, comprising 2 known and 11 unknown attacks, incorporating codec and transmission variability. The DF dataset consisted of 533,928 utterances, comprising 100 different spoofing attack algorithms, reflecting audio compression variability. To assess the generalizability, we also evaluated the proposed method on the In-The-Wild dataset~\cite{muller22_interspeech}, which comprised 11,816 spoofed utterances and 19,963 bona fide utterances, collected from podcasts and speeches.

We evaluated the model performance based on equal error rate (EER) across all datasets. For the LA dataset, we also computed the minimum tandem detection cost function (min t-DCF), to assess the impact of spoofing on speaker verification.

\subsection{Implementation details}
We adopted the Conformer-based model~\cite{rosello23_interspeech} as the baseline speech deepfake detection (SDD) model to evaluate the proposed methods. For feature extraction, we employed the pre-trained XLS-R 300M model\footnote{\url{https://github.com/pytorch/fairseq/tree/main/examples/wav2vec/xlsr}} \cite{babu22_interspeech} as the front-end, which takes raw waveforms as input and is jointly fine-tuned with the back-end. Each waveform was either truncated or concatenated to form segments of approximately 4 seconds, yielding 64,600 samples at a 16,000 Hz sampling rate.

All models were trained using the Adam optimizer with an initial learning rate of $10^{-6}$, a weight decay of $10^{-4}$, and a batch size of 20. A weighted cross-entropy loss was employed to mitigate class imbalance. To incorporate curriculum learning (CL), we designed the difficulty levels $H = \{0.35, 0.5, 0.65, 0.8, 1.0\}$ and the corresponding pacing sequence $T = \{1, 9, 17, 21, 23\}$. The total number of training epochs was set to 100. Early stopping was applied with a patience of seven epochs only when the difficulty level reached $1.0$. Additionally, for the dynamic temperature (DT), the MOS threshold \( m_{\text{th}} \) was set to $3.584$, yielding an error rate of $9.06\%$ when the training set was evaluated using MOS predictions. Dynamic temperature scaling was activated when the difficulty level reached 0.8. The final results were reported based on a weighted average of the five best-performing models on the validation set.

For comparison with other models, we applied RawBoost \cite{tak2022rawboost} data augmentation to all the models. Two different RawBoost algorithms were applied depending on the dataset. For the LA evaluation, we applied data augmentation with a combination of linear and non-linear convolutive noise and impulsive signal-dependent additive noise. For the DF evaluation, we applied augmentation with stationary signal-independent additive noise. For evaluations on the In-The-Wild dataset, we used models trained with the augmentation method for the LA dataset. The results are reproducible using our open-source code\footnote{\url{https://github.com/rlataewoo/nacl_sdd}}.

\vspace{-0.2cm}
\subsection{Results}

\begin{table}[t]
\centering
\caption{Comparison of EER (\%) between the proposed model and SOTA models on the ASVspoof 2021 evaluation sets.}
\vspace{-0.3cm}
\label{tab:SOTA_comparison}
\footnotesize
\begin{tabular}{l c c}
\toprule
\textbf{Model} & \textbf{LA} & \textbf{DF} \\  \midrule
XLS-R AASIST$^\dagger$~\cite{tak22_odyssey} & 1.00 & 3.69 \\
XLS-R AASIST2~\cite{zhang2024improving} & 1.61 & 2.77 \\
XLS-R Conformer~\cite{rosello23_interspeech} & 0.97 & 2.58 \\
WavLM MFA~\cite{guo2024audio} & 5.08 & 2.56 \\
XLS-R MoE~\cite{wang2024mixture} & 2.96 & 2.54 \\ 
XLS-R OC-ACS~\cite{kim24b_interspeech} & 1.30 & 2.19 \\ 
XLS-R SLS$^\dagger$~\cite{zhang2024audio} & 3.88 & 2.09 \\ 
XLS-R Conformer + TCM~\cite{truong24b_interspeech} & 1.03 & 2.06 \\ \midrule
XLS-R Conformer + CL \& DT & \textbf{0.89} & \textbf{1.88} \\ \bottomrule
\end{tabular}

\noindent
\footnotesize{$^\dagger$ The average result from different training sessions.}

  \vspace{-0.3cm}
\end{table}

\begin{table}[t]
  \centering
  \caption{Comparison between the proposed model and SOTA models on the In-The-Wild dataset.}
  \label{tab:InTheWild_comparison}
  \vspace{-0.2cm}
  \footnotesize
  \begin{tabular}{lr}
    \toprule
    \textbf{Model} & \textbf{EER(\%)} \\
    \midrule
    XLS-R AASIST~\cite{tak22_odyssey} (Reported by \cite{zhang2024audio}) & 10.46 \\
    XLS-R MoE~\cite{wang2024mixture} & 9.17  \\
    XLS-R SLS$^\dagger$~\cite{zhang2024audio}  & 8.87  \\
    \midrule
    XLS-R Conformer   & 7.29  \\
    XLS-R Conformer + CL    & \underline{6.83}  \\
    XLS-R Conformer + CL \& DT & \textbf{6.60} \\
    \bottomrule
  \end{tabular}
  
  \noindent
  \footnotesize{$^\dagger$ The average result from different training sessions.}
  
  \vspace{-0.6cm}
\end{table}

\begin{table}[t]
  \caption{Comparison of EER (\%) using different MOS predictors for our proposed method on the ASVspoof 2021 evaluation sets.}
  \label{tab:MOS_comparison}
  \vspace{-0.4cm}
  \footnotesize
  \centering
  \begin{tabular}{lccc}
    \toprule
    \textbf{MOS} & \textbf{Type}  & \textbf{LA} & \textbf{DF}              \\
    \midrule
    None      & -    & 1.09    & 2.45    \\ 
    DNSMOS        & Audio quality    & 1.57     & 2.57    \\
    NISQA-TTS      & Naturalness    & \underline{1.05}     & \underline{2.37}    \\ \midrule
    UTMOS (Ours)  & Naturalness    & \textbf{0.89}     & \textbf{1.88}    \\

    \bottomrule
  \end{tabular}
\vspace{-0.2cm}
\end{table}

\begin{table}[t]
  \caption{Ablation study on temperature scaling for the ASVspoof 2021 evaluation sets.}
  \label{tab:ablation_temperature}
  \vspace{-0.2cm}
  \footnotesize
  \centering
  \begin{tabular}{lccc}
    \toprule
    \textbf{Temperature}  & \textbf{LA} & \textbf{DF} \\
    \midrule
    $\tau=4$      & 1.03    & 5.37     \\ 
    $\tau=2$     & \textbf{0.84}    & 2.66   \\ 
    $\tau=1$      & 0.92    & 2.11   \\ 
    $\tau=1/2$      & 1.05    & \underline{2.02}    \\ 
    $\tau=1/4$      & 1.49    & 2.39   \\ \midrule
    DT (Ours) &    \underline{0.89}     & \textbf{1.88}  \\

    \bottomrule
  \end{tabular}
\vspace{-0.5cm}
\end{table}

\subsubsection{Assessing the impact of the proposed training strategy}
To assess the effectiveness of the proposed training strategy, we compared the models trained with and without curriculum learning (CL) and dynamic temperature (DT). Table~\ref{tab:proposed_performance} presents the results for three architectures: XLS-R AASIST~\cite{tak22_odyssey}, XLS-R Transformer, and XLS-R Conformer~\cite{rosello23_interspeech}, evaluated under fixed and variable utterance conditions for both the LA and DF. 

The results showed that when CL and DT were applied individually, all models, except for the XLS-R AASIST + DT, which experienced a slight performance drop for LA, exhibited an average improvement in performance. This indicates that both techniques independently contributed to a certain level of performance enhancement. Furthermore, integrating CL and DT led to an average improvement in detection performance across all architectures. Notably, the XLS-R Conformer achieved the lowest EER of 0.89\% for LA (Fix) and 1.88\% for DF (Fix), representing improvements of 18\% and 23\%, respectively, compared to the baseline XLS-R Conformer. These findings confirm that the proposed training strategy significantly enhances detection performance without requiring modifications to the model architecture, with the most substantial improvements observed in the XLS-R Conformer.

\vspace{-0.2cm}
\subsubsection{Comparison of the proposed model with SOTA models}

Table~\ref{tab:SOTA_comparison} presents a comparison between the proposed model and state-of-the-art (SOTA) models on the ASVspoof 2021 evaluation set. The proposed model achieved superior performance in both the LA and DF datasets. While the XLS-R Conformer + TCM improved DF detection, reducing the EER by 0.52\% compared to the XLS-R Conformer, it performed 0.06\% worse on the LA dataset. In contrast, the proposed training strategy achieved the lowest EER across both datasets without any modifications to the XLS-R Conformer architecture, demonstrating its effectiveness in improving detection performance.

To further evaluate the generalization performance, we conducted additional experiments using the In-The-Wild dataset. As shown in Table~\ref{tab:InTheWild_comparison}, the proposed model, enhanced with CL and DT, achieved the lowest EER of 6.60\%. Additionally, even without applying DT, it achieved an EER of 6.83\%, confirming that DT contributes to improving the generalization performance. These results demonstrate the robustness of the proposed model in effectively detecting spoofed speech under various conditions.

\subsubsection{Comparison of different MOS predictors for SDD}
Table~\ref{tab:MOS_comparison} presents a comparison of the performance of the proposed method using different automatic MOS prediction models. UTMOS and NISQA-TTS \cite{mittag20_interspeech} outperformed the baseline, demonstrating that using MOS predictors focused on naturalness leads to better detection performance. Conversely, DNSMOS \cite{reddy2021dnsmos}, which evaluates audio quality with a focus on denoising, resulted in decreased detection performance. This indicates that the spoofed speech is not easily distinguishable from bona fide speech in terms of audio quality. Therefore, considering naturalness as an additional cue in SDD proved to be more effective, supporting our hypothesis.

\subsubsection{Ablation study on temperature scaling}
Table~\ref{tab:ablation_temperature} presents an ablation study comparing the fixed temperature settings with the proposed dynamic temperature scaling. While certain fixed temperatures produced competitive results, they exhibited trade-offs between LA and DF performances. For instance, \( \tau = 2 \) achieved the best performance in LA but significantly degraded the DF performance. Similarly, \( \tau = 1/2 \) further enhanced the DF detection, resulting in degraded LA performance. These findings highlight the limitations of using a single fixed temperature for all samples. In contrast, the dynamic temperature scaling method achieved a balanced performance across LA and DF, demonstrating its effectiveness in improving generalizability.

\section{Conclusion}
In this study, we introduce a naturalness-aware training strategy that combines curriculum learning and dynamic temperature scaling to enhance speech deepfake detection performance. We present an efficient learning method that leverages perceptual quality, particularly naturalness, to measure sample difficulty and progressively introduce more challenging training samples, while adjusting dynamic temperature to lower the confidence of difficult samples and increase the confidence of easy samples. Experiments conducted on the ASVspoof 2021 and In-The-Wild datasets demonstrate that the proposed approach significantly improves detection performance without modifying the model architecture. These results suggest that integrating naturalness into the training process is essential for robust and generalizable deepfake detection.

\section{Acknowledgements}
\vspace{-0.25cm}
This work was partly supported by Institute of Information \& communications Technology Planning \& Evaluation (IITP) grant funded by the Korea government (MSIT) (No.2022-0-00963 and No.RS-2024-00456709).
\vspace{-0.45cm}

\bibliographystyle{IEEEtran}
\bibliography{main}

\end{document}